\documentclass[conference]{IEEEtran}
\IEEEoverridecommandlockouts
\usepackage{cite}
\usepackage{amsmath,amssymb,amsfonts}
\usepackage{algorithmic}
\usepackage{graphicx}
\usepackage{textcomp}
\usepackage{xcolor}
\usepackage{hyperref}
\usepackage{braket}
\usepackage{amsthm}

\providecommand{\myvec}[1]{\ensuremath{\boldsymbol{#1}}}

\providecommand{\tt}{\ensuremath{\myvec{t}}}

\newtheorem{theorem}{Theorem}

\newtheorem{definition}[theorem]{Definition}

\def\BibTeX{{\rm B\kern-.05em{\sc i\kern-.025em b}\kern-.08em
    T\kern-.1667em\lower.7ex\hbox{E}\kern-.125emX}}

\renewcommand{\figureautorefname}{Figure~\negthinspace}
\renewcommand{\equationautorefname}{Equation~\negthinspace}

\begin{document}

\title{Learning to Measure Quantum Neural Networks
\thanks{The views expressed in this article are those of the authors and do not represent the views of Wells Fargo. This article is for informational purposes only. Nothing contained in this article should be construed as investment advice. Wells Fargo makes no express or implied warranties and expressly disclaims all legal, tax, and accounting implications related to this article.}
}

\author{
\IEEEauthorblockN{Samuel Yen-Chi~Chen}
\IEEEauthorblockA{\textit{Wells Fargo} \\
New York NY, USA \\
yen-chi.chen@wellsfargo.com}
\and 
\IEEEauthorblockN{Huan-Hsin~Tseng}
\IEEEauthorblockA{\textit{AI Department} \\
\textit{Brookhaven National Laboratory}\\
Upton NY, USA  \\
htseng@bnl.gov}
\and
\IEEEauthorblockN{Hsin-Yi~Lin}
\IEEEauthorblockA{\textit{Department of Mathematics} \\
\textit{and Computer Science} \\
\textit{Seton Hall University}\\
South Orange NJ, USA \\
hsinyi.lin@shu.edu}
\and
\IEEEauthorblockN{Shinjae~Yoo}
\IEEEauthorblockA{\textit{AI Department} \\
\textit{Brookhaven National Laboratory}\\
Upton NY, USA \\
syjoo@bnl.gov}
}

\maketitle

\begin{abstract}
The rapid progress in quantum computing (QC) and machine learning (ML) has attracted growing attention, prompting extensive research into quantum machine learning (QML) algorithms to solve diverse and complex problems.
Designing high-performance QML models demands expert-level proficiency, which remains a significant obstacle to the broader adoption of QML. A few major hurdles include crafting effective data encoding techniques and parameterized quantum circuits, both of which are crucial to the performance of QML models. Additionally, the measurement phase is frequently overlooked—most current QML models rely on pre-defined measurement protocols that often fail to account for the specific problem being addressed.
We introduce a novel approach that makes the observable of the quantum system—specifically, the Hermitian matrix—learnable. Our method features an end-to-end differentiable learning framework, where the parameterized observable is trained alongside the ordinary quantum circuit parameters simultaneously.
Using numerical simulations, we show that the proposed method can identify observables for variational quantum circuits that lead to improved outcomes, such as higher classification accuracy, thereby boosting the overall performance of QML models.
\end{abstract}


\section{Introduction}
\begin{figure}[htbp]
\begin{center}
\includegraphics[width=1\columnwidth]{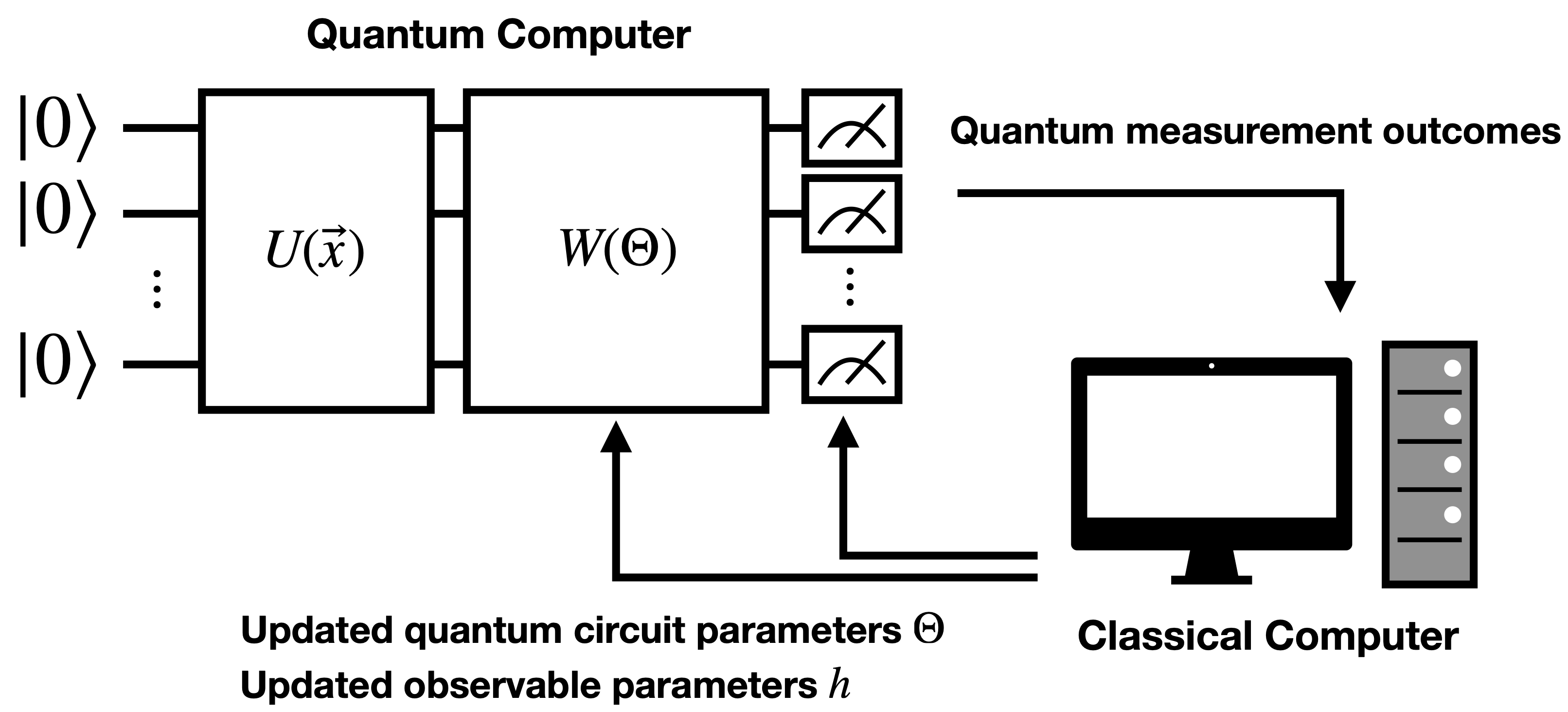}
\caption{{\bfseries Hybrid Quantum-Classical Computing with Learnable Measurements.}}
\label{fig:hybrid_quantum_classical_scheme}
\end{center}
\vskip -0.15in
\end{figure}

The intersection of quantum computing (QC) and machine learning (ML) has garnered significant attention in recent years, driven by advancements in both quantum hardware and AI/ML technologies. Quantum machine learning (QML) is an emerging field that leverages the principles of quantum mechanics to enhance the performance of machine learning models. Despite the limitations of current quantum computers, hybrid quantum-classical algorithms have been developed to exploit the strengths of both computing paradigms.
Variational Quantum Algorithms (VQAs) \cite{bharti2022noisy} represent a class of hybrid quantum-classical algorithms where quantum circuit parameters are optimized via classical methods such as gradient descent \cite{mitarai2018quantum} or metaheuristic techniques like evolutionary algorithms \cite{chen2022variationalQRL}. These algorithms enable quantum circuit models, such as Quantum Neural Networks (QNNs) or Variational Quantum Circuits (VQCs), to address a wide range of AI/ML tasks, including classification \cite{chen2021end, qi2023qtnvqc, mitarai2018quantum, chen2022quantumCNN,chen2024qeegnet,lin2024quantumGRADCAM}, time-series forecasting \cite{chen2022quantumLSTM,chehimi2024FedQLSTM,chen2024QFWP}, natural language processing \cite{li2023pqlm, yang2022bert, di2022dawn, stein2023applying}, reinforcement learning \cite{chen2020VQDQN, chen2022variationalQRL, chen2023QLSTM_RL, skolik2022quantum, jerbi2021parametrized,yun2022quantum} and model compression \cite{liu2024training,liu2024qtrl,liu2024federated_QT,lin2024_QT_DeepFake,lin2024QT_LSTM}.
Ordinary QNNs are trained with pre-defined observables $\hat{B}_{k}$ such as the Pauli-$X$, $Y$ and $Z$ observables. However, the conventional choice of Pauli matrix $X, Y, Z$ has only eigenvalues $\lambda = \pm 1$ so that the VQC prediction is always confined in $\bra{\psi} H \ket{\psi} \in [-1, 1]$ regardless of the unitary gate $U(\vec{x}), W(\Theta)$ used, which poses a restriction on VQC capabilities. By Rayleigh quotient, we know that $ \lambda_{\text{min}} \leq \bra{\psi} H \ket{\psi} \leq \lambda_{\text{max}}$ for any normalized wave function $\| \psi \| =1$. Therefore, increasing the range of the Hermitian spectrum (eigenvalues) will increase the VQC output range for versatile ML tasks, such as classifications and regressions. 
In this paper, we propose an approach to automatically discover observables through end-to-end gradient-based optimization, as illustrated in \figureautorefname{\ref{fig:hybrid_quantum_classical_scheme}}. Specifically, we parameterize the Hermitian matrix used as observables and train these parameters concurrently with the standard quantum circuit parameters (rotation angles). Numerical simulations demonstrate that the proposed learnable observable framework outperforms standard VQC training with fixed observables.
\section{Related Work}
Developing a successful QML model for a specific task demands expert knowledge in quantum information science. To expand the range of QML applications, significant efforts have been dedicated to creating automated procedures for designing quantum circuit models. These approaches are explored within the field of Quantum Architecture Search (QAS) \cite{martyniuk2024quantum}, which leverages various search algorithms and machine learning techniques to generate high-performance quantum architectures.
Several approaches, such as reinforcement learning (RL) \cite{kuo2021quantum,ye2021quantum,zhu2023quantum,sogabe2022model,wang2024rnn,fodera2024reinforcement,chen2023quantumRL_QAS,dai2024quantum} and evolutionary algorithms \cite{ding2022evolutionary,chen2024EvoQAS_ED}, have been explored to address the challenges in QAS. However, a significant challenge with both RL and evolutionary algorithms is the need to assess circuit performance each time a specific architecture is selected. This evaluation becomes increasingly difficult as the search space expands, for example, with the addition of more qubits or circuit components. Furthermore, these methods require tuning multiple hyperparameters, such as mutation rates and crossover probabilities for evolutionary algorithms, and exploration-exploitation ratios for RL techniques. Differentiable search has emerged as an alternative approach to mitigate these challenges.
Differentiable programming has been utilized in QAS for a range of QML tasks, including optimization, classification, and reinforcement learning \cite{zhang2022differentiable,sun2023differentiable,chen2024differentiable}. A key advantage of differentiable QAS is that the parameters governing the quantum circuit architectures are optimized concurrently with the parameters (rotation angles) of the VQCs. Additionally, differentiable search requires fewer hyperparameters compared to evolutionary or RL-based approaches, simplifying the optimization process.
This paper distinguishes itself from prior work by applying differentiable programming to optimize the measurement process of the quantum system, rather than focusing on the optimization of the VQC architecture itself. Unlike the approach in \cite{yun2022slimmable}, our method allows for the training of any number of observables or Hermitian matrices for systems with any number of qubits. Additionally, the parameters of the observables are optimized simultaneously with the standard rotation angles, with the flexibility to employ separate optimizers for each.
\section{Quantum Neural Networks}
\begin{figure}[htbp]
\vskip -0.15in
\begin{center}
\includegraphics[width=1\columnwidth]{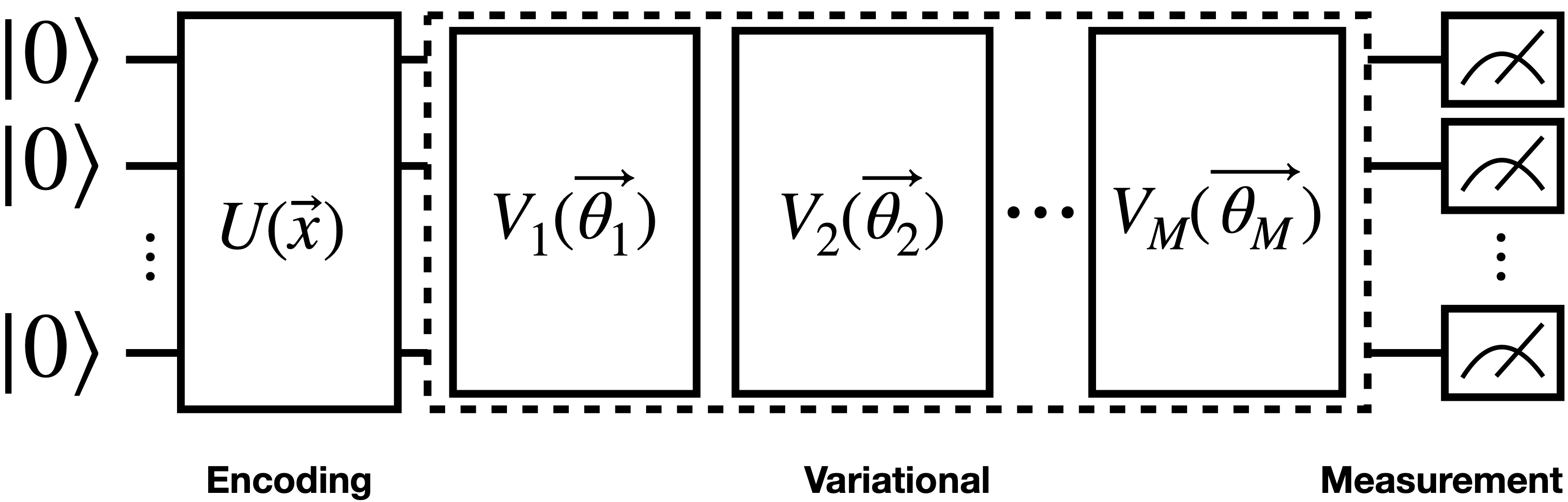}
\caption{{\bfseries Generic Structure of a Variational Quantum Circuit (VQC).}}
\label{fig:vqc_scheme}
\end{center}
\vskip -0.15in
\end{figure}

A VQC, also referred to as a Parameterized Quantum Circuit (PQC), generally comprises three key components: the \emph{encoding circuit}, the \emph{variational or parameterized circuit}, and the final \emph{quantum measurement}.
The encoding circuit, denoted as $ U(\vec{x})$, is designed to map the input vector of classical numerical values $\vec{x}$ into a quantum state, transforming it into $U(\vec{x})\ket{0}^{\otimes n}$, where $\ket{0}^{\otimes n}$ represents the ground state of the quantum system and $n$ denotes the number of qubits.
The parameterized circuit then processes and transforms the encoded state and it becomes $W(\Theta)U(\vec{x})\ket{0}^{\otimes n}$. Generally, the variational (parameterized or learnable) circuit $W(\Theta)$ is constructed by multiple layers of trainable circuit layer $V_{j}(\vec{\theta_{j}})$ (illustrated in \figureautorefname{\ref{fig:vqc_scheme}}), denoted as $W(\Theta) = \prod_{j = M}^{1} V_{j}(\vec{\theta_{j}})$, where $\Theta$ represents the collection of all learnable parameters $\{\vec{\theta_{1}} \cdots \vec{\theta_{M}}\}$. Then the quantum state vector generated by the encoding circuit and variational circuit can be shown as, 
\begin{equation}
\label{eqn:vqc_state_psi}
    \ket{\Psi} = W(\Theta) U(\vec{x})\ket{0}^{\otimes n} = \left( \prod_{j = M}^{1} V_{j}(\vec{\theta_{j}}) \right) U(\vec{x})\ket{0}^{\otimes n}
\end{equation}
Information from the VQC can be extracted by performing measurements using predefined observables, denoted as $\hat{B}_{k}$. The VQC operation can be seen as a quantum function $\overrightarrow{f(\vec{x} ; \vec{\theta})}=\left(\left\langle\hat{B}_1\right\rangle, \cdots,\left\langle\hat{B}_n\right\rangle\right)$, where $\left\langle\hat{B}_{k}\right\rangle =\left\langle 0\left|U^{\dagger}(\vec{x})W^{\dagger}(\Theta) \hat{B}_{k} W(\Theta)U(\vec{x})\right| 0\right\rangle$. Expectation values $\left\langle\hat{B}_{k}\right\rangle$ can be estimated by conducting repeated measurements (shots) on physical quantum devices or by direct calculation when employing quantum simulation tools. Usually, the observable $\hat{B}_{k}$ is a predefined Hermitian matrix. A common choice is the Pauli-$Z$ matrix.
\section{Learnable Quantum Measurements}
\begin{definition}
A Hermitian matrix $B$ is used as an \emph{observable} in quantum mechanics, for it yields real-valued expectations $\bra{\Psi} B \ket{\Psi}$ under a wave function $\ket{\Psi}$. The condition $B = B^{\dagger}$ requires each matrix element $b_{ij} = \overline{b_{ji}} \in \mathbb{C}$. Consequently, a Hermitian matrix can be generated by $2 \times \frac{N(N-1)}{2} + N = N^2$ real parameters,
\begin{equation}
    B = \begin{pmatrix}
d_{11} & a_{12} + i c_{12} & a_{13} + i c_{13} & \cdots & a_{1N} + i c_{1N}  \\
* & d_{22}  & a_{23} + i c_{23}  & \cdots & a_{2N} + i c_{2N}  \\
* & * & d_{33}  & \cdots & a_{3N} + i c_{3N}  \\
\vdots & \vdots & \vdots & \ddots & \vdots \\
* & * & * & \cdots & d_{NN}
\end{pmatrix}
\nonumber
\end{equation}
where $*$ denotes the corresponding complex conjugate and $a_{ij}$, $c_{ij}, d_{ii}$ are arbitrary real numbers.
\end{definition}
The Hermitian matrix can be initialized randomly and optimized iteratively via gradient-based methods. To elucidate the process, we can write the Hermitian matrix of a $n$-qubit system with $N = 2^n$ as the parametrization of coefficient $\vec{b} = (b_{11}, \ldots, b_{NN}) \in \mathbb{C}^{N \times N}$ as $B(\vec{b}) = \sum_{i = 1}^N\sum_{j = 1}^N b_{ij} \, E_{ij}$, where $b_{ij} = \overline{b_{ji}}$ and $E_{ij}$ as the indicating matrix with only one non-zero at entry $(i,j)$, 
\begin{equation}
E_{ij} = 
\begin{pmatrix}
0 & 0 & \cdots & 0 & 0 \\
0 & 0 & \cdots & 0 & 0 \\
\vdots  & \vdots & \ddots & \vdots & \vdots \\
0 & 0 & \cdots & 1 & 0 \\
0 & 0 & \cdots & 0 & 0
\end{pmatrix}
\end{equation}
Given a quantum state $\ket{\Psi}$ (e.g. the one shown in \equationautorefname{\ref{eqn:vqc_state_psi}}), the expectation value with the $B(\vec{b})$ can be written as,
\begin{align}
    \bra{\Psi}B(\vec{b})\ket{\Psi} 
    &= \bra{\Psi}\sum_{i}^{N} \sum_{j}^{N} b_{ij} E_{ij}\ket{\Psi} \\
    &= \sum_{i}^{N} \sum_{j}^{N} b_{ij}  \bra{\Psi}E_{ij}\ket{\Psi}
\end{align}
The total differential $\bra{\Psi}B(\vec{b})\ket{\Psi}$ is 

\begin{equation}
    \nabla \bra{\Psi}B(\vec{b})\ket{\Psi} = 
\begin{pmatrix}
    &\nabla_{\theta} \bra{\Psi}B(\vec{b})\ket{\Psi}\\ &\nabla_{\vec{b}} \bra{\Psi}B(\vec{b})\ket{\Psi}
\end{pmatrix}
\end{equation}
where the differentiation of $\bra{\Psi}B(\vec{b})\ket{\Psi}$ can be written as,
\begin{align}
    &\frac{\partial \bra{\Psi} B(\vec{b}) \ket{\Psi}}{\partial b_{k\ell}}
    =\frac{\partial \bra{0}U(\vec{x})^\dagger W(\Theta)^\dagger B(\vec{h}) W(\Theta) U(\vec{x})\ket{0}}{\partial b_{k\ell}} \nonumber  \\
    &= \frac{\partial}{ \partial b_{k\ell}} \sum_{i}^{N} \sum_{j}^{N} b_{ij}  \bra{0}U(\vec{x})^\dagger W(\Theta)^\dagger E_{ij} W(\Theta) U(\vec{x})\ket{0} \nonumber \\
    &=  \sum_{i}^{N} \sum_{j}^{N} \delta_{ik}\delta_{j \ell}  \bra{0}U(\vec{x})^\dagger W(\Theta)^\dagger E_{ij} W(\Theta) U(\vec{x})\ket{0} \nonumber  \\
    &= \bra{0}U(\vec{x})^\dagger W(\Theta)^\dagger E_{k \ell }W(\Theta) U(\vec{x})\ket{0}  \nonumber \\
    &= \overline{ \big( W(\Theta) U(\vec{x}) )}_{k 1} \big( W(\Theta) U(\vec{x}) \big)_{\ell 1} \nonumber
\end{align}
where the bar is the complex conjugate and $k, \ell \in \{1, \ldots, N\}$ denote matrix indices. $\Theta$ represents the parameters of the variational quantum circuit. The last equality shows the explicit dependency on $W(\Theta)$.
\section{Experiments}
\subsection{Task 1: Classification}
We begin by evaluating the proposed methods on the standard \texttt{make\_moons} dataset from \texttt{scikit-learn}. The dataset is generated with noise levels of 0.1, 0.2, and 0.3. The training settings are as follows: batch size of 20, training set size of 200, testing set size of 100, 4 qubits, 2 layers in the VQC trainable circuit, a learning rate of $1 \times 10^{-2}$, and a learning rate of $1 \times 10^{-1}$ for optimizing the Hamiltonian $H$. 
The optimizer used for both the standard VQC and the VQC with learnable observables is RMSProp. In the case where the VQC and learnable observables utilize separate optimizers, RMSProp is applied to the VQC parameters, while Adam is used for the parameters of the observables.
For the learnable observable component, each class introduces $4$ additional trainable parameters. For instance, in the \texttt{make\_moons} dataset used for binary classification, there are $8$ parameters dedicated to the learnable observable. 
We repeat the experiments five times to calculate the mean and standard deviation, providing a comprehensive evaluation of the model’s performance.
\begin{figure}[htbp]
\begin{center}
\includegraphics[width=1\columnwidth]{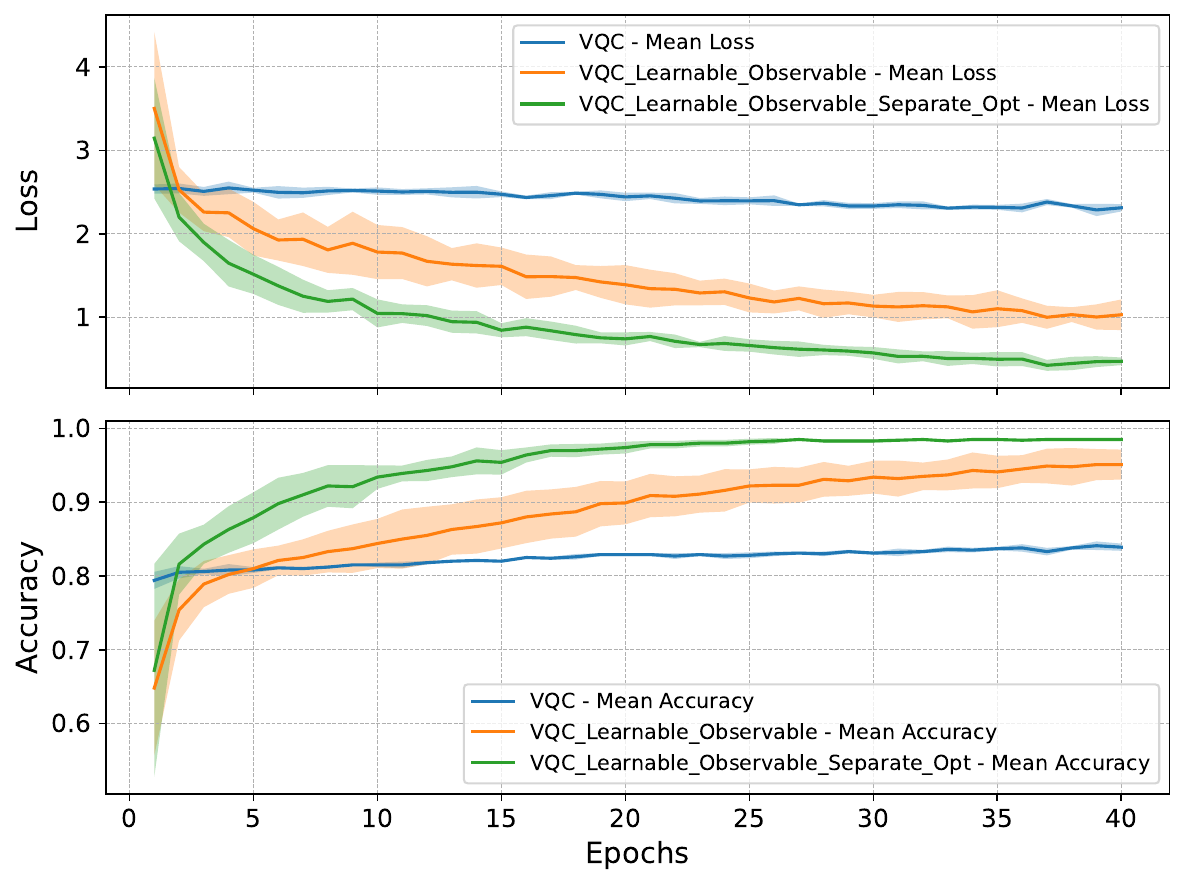}
\caption{{\bfseries Comparison of different VQC models in \texttt{make\_moons} dataset with noise = 0.1.}}
\label{fig:results_make_moons_noise_0.1}
\end{center}
\end{figure}
\begin{figure}[htbp]
\begin{center}
\includegraphics[width=1\columnwidth]{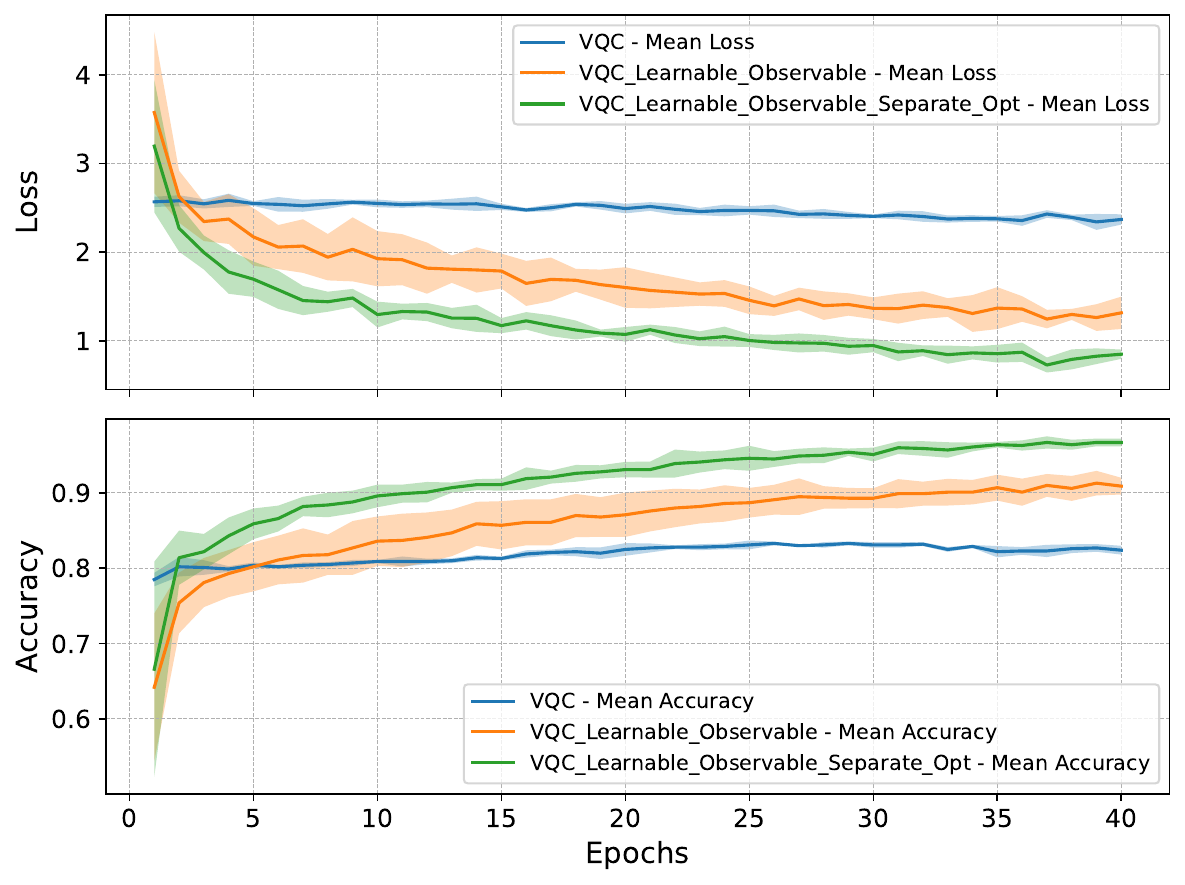}
\caption{{\bfseries Comparison of different VQC models in \texttt{make\_moons} dataset with noise = 0.2.}}
\label{fig:results_make_moons_noise_0.2}
\end{center}
\end{figure}
\begin{figure}[htbp]
\begin{center}
\includegraphics[width=1\columnwidth]{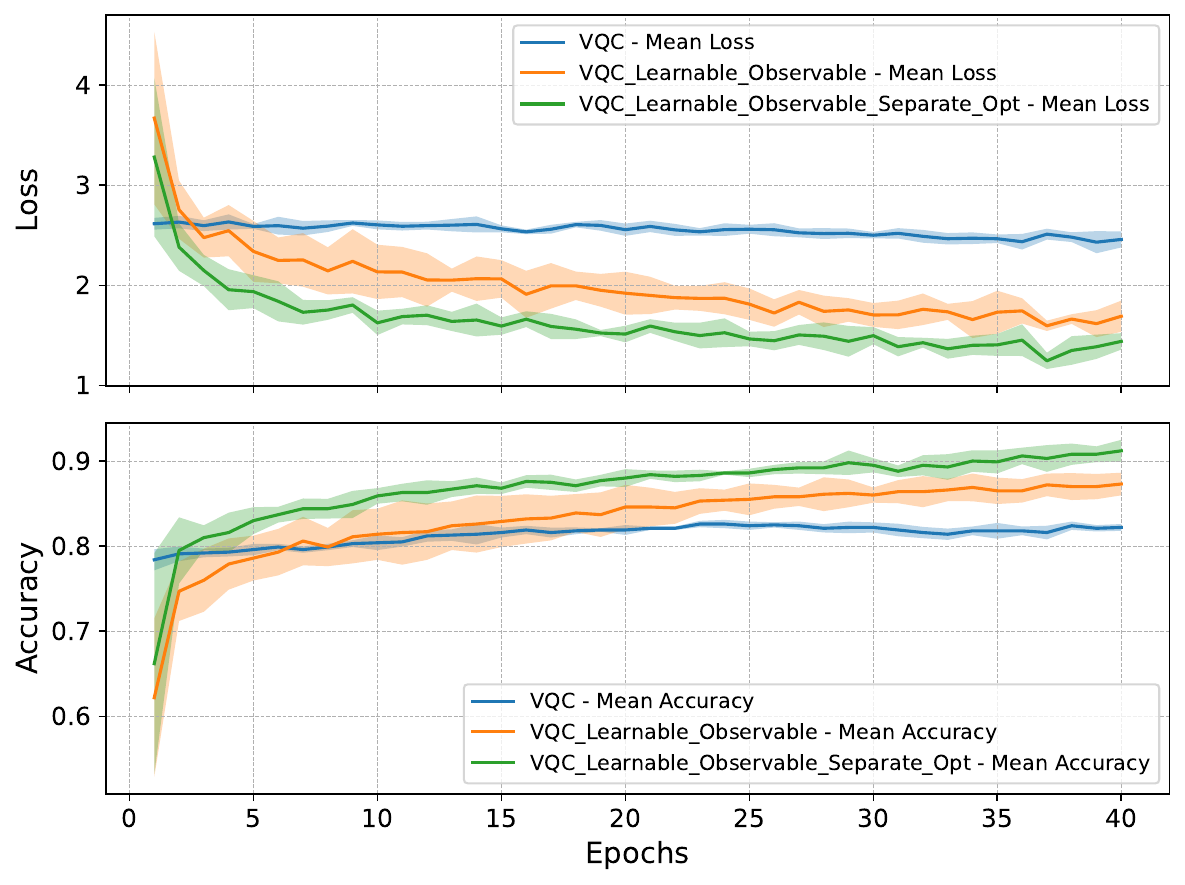}
\caption{{\bfseries Comparison of different VQC models in \texttt{make\_moons} dataset with noise = 0.3.}}
\label{fig:results_make_moons_noise_0.3}
\end{center}
\end{figure}
As shown in \figureautorefname{\ref{fig:results_make_moons_noise_0.1}}, with noise set to $0.1$ in the \texttt{make\_moons} dataset, the VQC model incorporating both a learnable observable and a separate optimizer (with distinct learning rates) outperforms the VQC model with only a learnable observable. The standard VQC model without learnable observables exhibits the lowest performance. We further increased the difficulty of the \texttt{make\_moons} dataset by raising the noise level to $0.2$ and $0.3$. As shown in \figureautorefname{\ref{fig:results_make_moons_noise_0.2}} and \figureautorefname{\ref{fig:results_make_moons_noise_0.3}}, the VQC model with both a learnable observable and a separate optimizer (with distinct learning rates) continues to outperform the VQC model with learnable observables trained with the same optimizer, while the conventional VQC model remains the lowest performer in these scenarios.

\begin{figure}[ht]
\begin{center}
\centerline{\includegraphics[width=1\columnwidth]{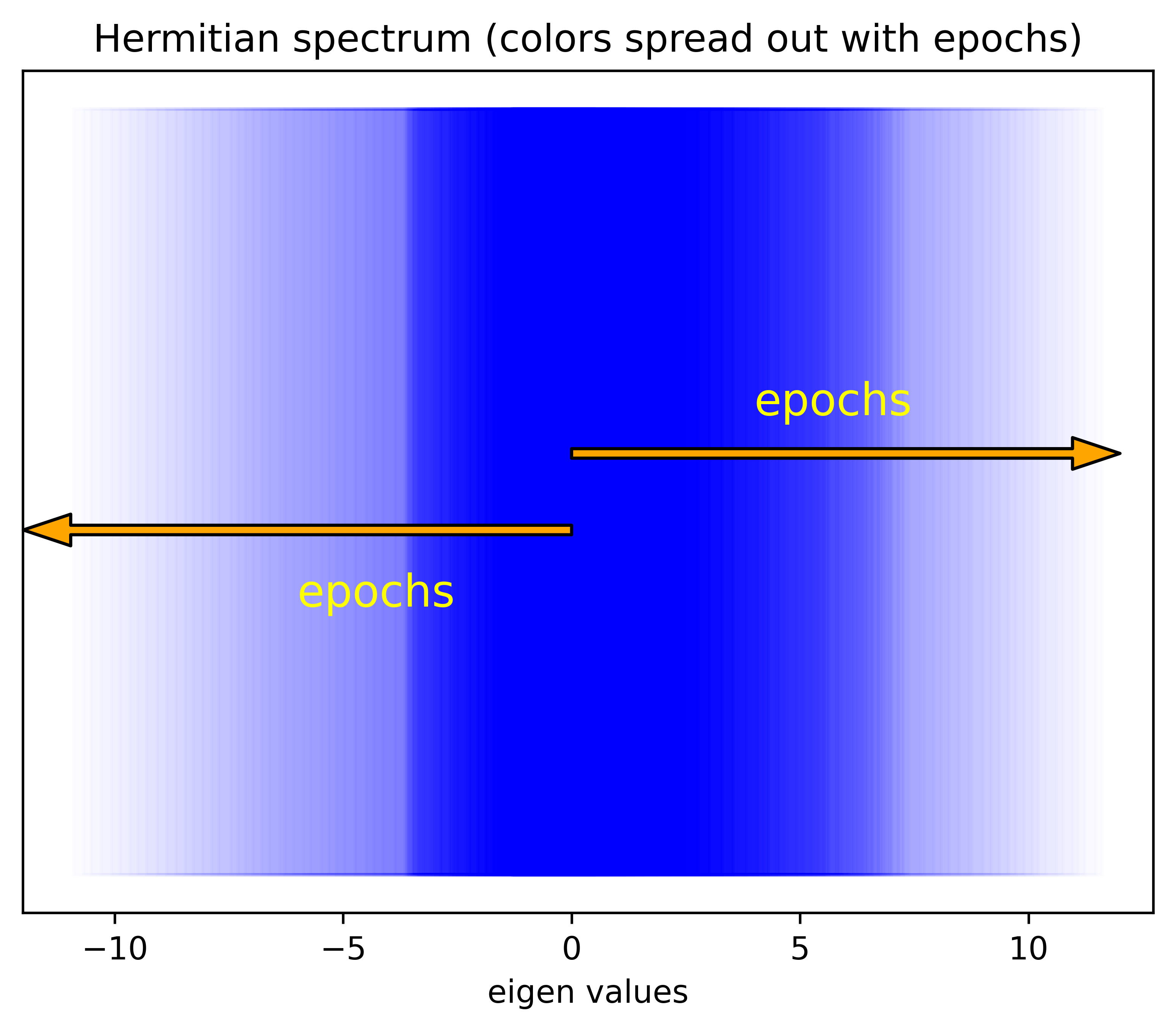}}
\caption{{\bfseries [Make moon experiment] Maximal and minimal eigenvalues of $H$ on each training epoch indicating the range is spreading out with the epochs.}}
\label{fig: spectral}
\end{center}
\vskip -0.15in
\end{figure}

\subsection{Task 2: Speaker Recognition}
To evaluate our method on more challenging tasks, we utilized a real-world dataset for speaker recognition based on acoustic speech signals. 

\paragraph{\textbf{Data}}
We adopted the \href{https://datashare.ed.ac.uk/handle/10283/2791}{Voice Bank corpus}~\cite{veaux2013voice} (VCTK) of 28 speakers~\cite{botinhao2016speech} (14 male, 14 female); each has approximately 400 \emph{clean} sentences. For our experiments, we selected 5 male and 5 female out of the 28 speakers and resampled the audio to 16 kHz. The data was split into training and testing sets with a 0.15 ratio, resulting in about 340 sentences for training and 60 for testing per speaker.

The waveforms were subsequently transformed into spectrograms via the Short-Time Fourier Transform (STFT), with amplitudes extracted as input features to form the final speaker recognition dataset of 10 speakers (classes). The resulting spectrograms have an input size of $(257, 128)$, where the first dimension 257 is due to the 128 Fourier frequency basis used while the second dimension 128 can be roughly regarded as the time evolution of the temporal sequence. The processed speech dataset is available here: \url{https://bnlbox.sdcc.bnl.gov/index.php/s/HwLAJWcqtntzayc}.

\paragraph{\textbf{Model architecture}}
A hybrid model joining classical networks with a VQC layer at the end was deployed. Three consecutive CNN layers were used to encode spectrograms regarded as 2D images into vectors of dimension 12,672. A linear layer was then employed to reduce the dimensionality to 10, which then served as the input to a VQC of 10 qubits. Eventually, the VQC with the structure of \figureautorefname{\ref{fig:vqc_scheme}} classifies speakers.

\begin{figure}[htbp]
\begin{center}
\includegraphics[width=1\columnwidth]{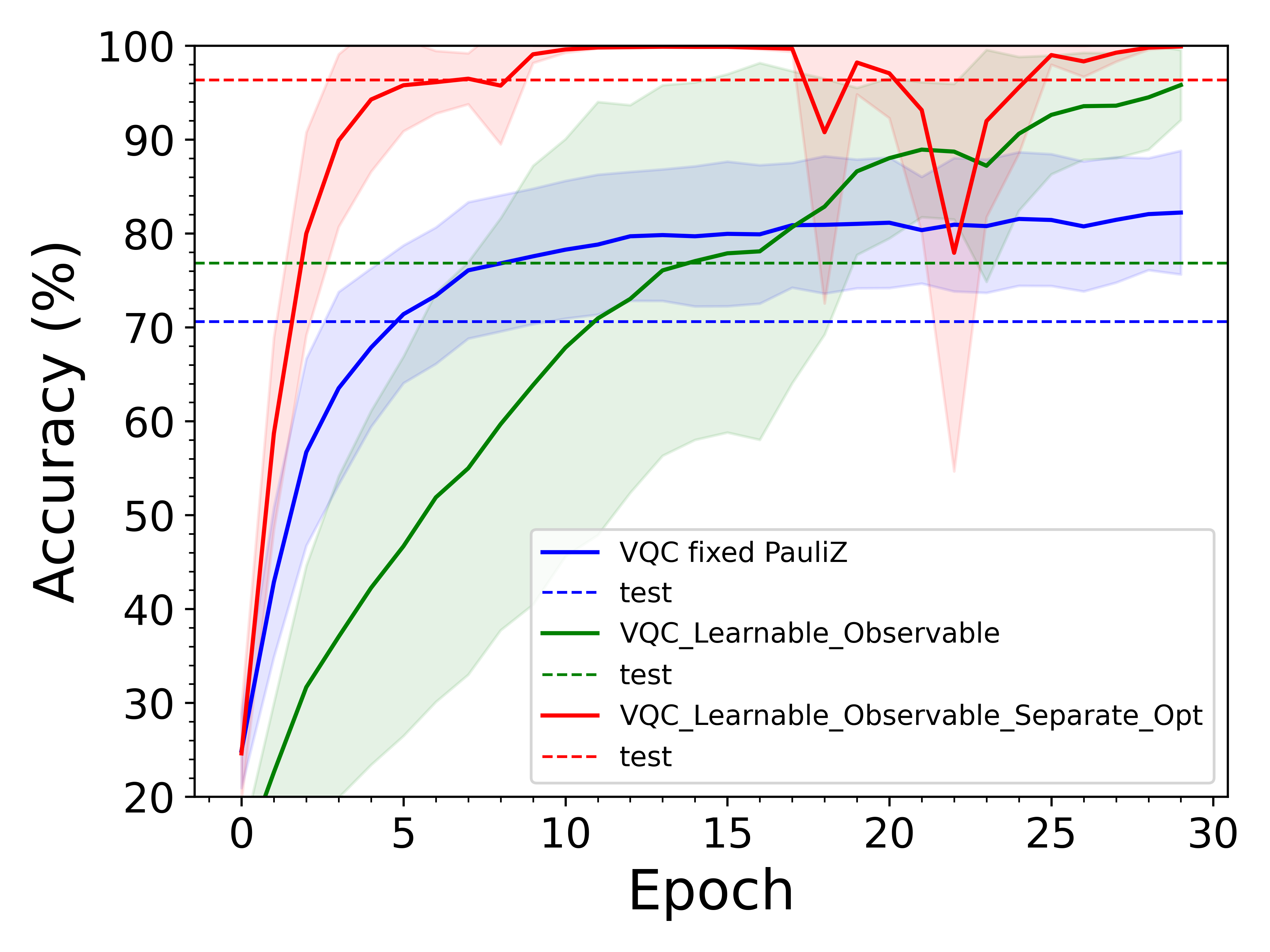}
\caption{{\bfseries Comparison of different VQC models on VCTK speaker recognition task.}}
\label{fig:results_VCTK}
\end{center}
\vskip -0.15in
\end{figure}

\paragraph{\textbf{Results}}
On this fixed model architecture, we analyze the effect of varying observables for the VQC performance. In \figureautorefname{\ref{fig:results_VCTK}}, experiments of 5 independent trials were shown; each contained 30 epochs of training on every model. The \emph{solid} lines are the mean of the training accuracies, and the \emph{dashed} lines are the final testing results on a test set disjoint from the training set. The shades enclosing the solid curves show the variance of the 5 trials.

It is observed that learning a proper Hermitian measurement rather than conventionally fixed Pauli matrices significantly raises performance. On the final test, the VQC with \emph{fixed} Pauli-$Z$ measurement (blue line) had accuracy $70.59\%$, while VQC of varying Hermitian (green line) attained $76.83\%$. Endowing different parameter learning rates to the unitary gates and the Hermitian (red line) further reached $96.33\%$. From the result, the reason for the optimizer separation is manifest as the qubit system demands large eigenvalue expansion for complex tasks so that the Hermitian measurement needs to grow quicker with a larger learning rate $0.1$ compared to that of the unitary gate $10^{-3}$. Other optimization techniques for further improvement are possible; here using different parameter optimizers serves to demonstrate the concept.

\section{Conclusion}
In this paper, we demonstrate that by making observables in QNN models learnable parameters to incorporate with variational unitary gates significantly enhances the performance of QNN-based machine learning tasks, such as classification and speech recognition. We show that the parameters governing quantum system measurements can be optimized along with the rotation angles in quantum gates, allowing the entire system to be trained end-to-end using differentiable methods. Consequently, we observe that the VQC tends to expand the range of the Hermitian eigenvalues whenever possible to achieve better predictions. 
Our approach opens the door to a more generalized framework for designing QNN-based models in AI and ML applications.
\clearpage
\bibliographystyle{IEEEtran}
\bibliography{bib/qml_examples,bib/vqc,bib/qas,bib/qt}
\end{document}